\documentclass[12pt]{article}
\newlength{\mytopmargin}
\newlength{\myleftmargin}
\usepackage{amsmath,amsthm,amssymb,amsbsy,epsfig,graphicx,color,multicol,subfigure}
\usepackage{array,calc}
\usepackage[enableskew]{youngtab}
\usepackage{rotating}
\usepackage{young,epic}
\usepackage{a4wide,bm}
\usepackage{url}
\usepackage{hyperref}

\newtheorem{conj}{Conjecture}
\newtheorem{prop}{Proposition}
\newtheorem{lemma}{Lemma}
\newtheorem{cor}{Corollary}

\usepackage{amsmath,amsfonts,amssymb}

\usepackage{graphicx}

\begin{document}
%

\title{Asymptotics of spacing distributions at the hard edge for $\beta$-ensembles}
\author{Peter J. Forrester }
\date{}
\maketitle
\noindent
\thanks{\small Department of Mathematics and Statistics, 
The University of Melbourne,
Victoria 3010, Australia email:  p.forrester@ms.unimelb.edu.au 
}

\begin{abstract}
\noindent In a previous work [J. Math. Phys. {\bf 35} (1994), 2539--2551], generalized hypergeometric functions have been used
to a give a rigorous derivation of  the large  $s$ asymptotic form of the general $\beta > 0$
gap probability $E_\beta^{\rm hard}(0;(0,s);\beta a/2)$, provided both
 $\beta a /2 \in \mathbb Z_\ge 0$ and
$2/\beta \in \mathbb Z^+$. It shown how the details of this method can be extended to remove
the requirement that $2/\beta \in \mathbb Z^+$. Furthermore, a large deviation formula for the gap
probability $E_\beta(n;(0,x);{\rm ME}_{\beta,N}(\lambda^{a \beta /2} e^{\beta N \lambda/2}))$ is
deduced by writing it in terms of the charateristic function
 of a certain linear statistic. By scaling $x = s/(4N)^2$ and taking $N \to \infty$, this is shown
 to reproduce a recent conjectured formula for  $E_\beta^{\rm hard}(n;(0,s);\beta a/2)$, $\beta a /2 \in \mathbb Z_{\ge 0}$,
 and moreover to give a prediction without the latter restriction.
 This extended formula, which for the constant term involves the Barnes double gamma function,
 is shown to satisfy  an asymptotic functional equation relating the
 gap probability with parameters $(\beta,n,a)$, to a gap probability with parameters $(4/\beta,n',a')$, where
 $n'=\beta(n+1)/2-1$, $a'=\beta(a-2)/2+2$.

\end{abstract}

\section{Introduction}
The topic of eigenvalue spacing distributions within random matrix theory is of interest both for its practical utility in comparisons with
experimental data, and for its rich mathematical content leading to many explicit functional forms. Moreover, the function theory associated
with the latter is intimately related to integrable systems theory, and this in turn offers a number of powerful asymptotic methods to further
quantify the spacing distributions. We refer to \cite{Fo05a,Bo09} for survey articles relating to explicit functional forms, and to \cite{FS12} for
a recent review on the associated asymptotics.

Spacing distributions depend on a symmetry parameter $\beta$, and the region of the eigenvalue spectrum being scaled. In  relation to $\beta$,
according to Dyson's three fold way \cite{Dy62c}, random matrix ensembles corresponding to quantum systems with a time reversal symmetry
$T$ such that $T^2= 1$ must have a probability density function invariant under conjugation by orthogonal matrices. If the quantum system does
not have a time reversal symmetry, or if the time reversal symmetry is such that $T^2 = -1$, the probability density function must  be invariant under
conjugation by unitary matrices, or by unitary symplectic matrices respectively. The three cases are labelled by the parameter $\beta$, with
$\beta = 1$ for orthogonal symmetry, $\beta = 2$ for unitary symmetry and $\beta = 4$ for unitary symplectic symmetry. One reason for this
labelling is that the joint  probability density function of $2 \times 2$ matrices with these symmetries is generically proportional to $|\lambda_1 -
\lambda_2|^\beta$ for near degenerate levels.

Regarding the region of the eigenvalue spectrum being scaled, depending on the details of the ensemble under consideration the
eigenvalue density $\rho_{(1)}(\lambda)$ can exhibit various asymptotic forms. Our interest is when
\begin{equation}\label{d1}
\rho_{(1)}(\lambda) = 0 \quad {\rm for} \quad \lambda < 0, \qquad
\rho_{(1)}(\lambda)  \sim
\displaystyle  {1 \over 2 \pi \lambda^{1/2} } \quad {\rm for} \quad \lambda \to \infty.
\end{equation}
This corresponds to the so-called hard edge scaled state. As a concrete example of a random matrix ensemble which exhibits this
behaviour, consider an $n \times p$ $(n \ge p)$ real Gaussian matrix $X$, each entry independently distributed as a standard normal,
and form the corresponding covariance matrix $X^\dagger X$. The joint eigenvalue probability density function is then proportional
to 
\begin{equation}\label{AB}
\prod_{l=1}^p \lambda_l^{a \beta /2} e^{- \beta \lambda_l/2} \prod_{1 \le j < k \le p} | \lambda_k - \lambda_j |^\beta, \qquad \lambda_l > 0,
\end{equation}
with $a = n - p + 2/\beta - 1$ and $\beta = 1$. The same construction, but with $X$ a standard complex Gaussian matrix gives (\ref{AB}) in the
case $\beta = 2$. The probability density function (\ref{AB}) for general parameters will be referred to as
${\rm ME}_{\beta,N}(\lambda^{a\beta/2} e^{-\beta \lambda/2})$. For general $\beta > 0$, fixed $a \ge 0$, and with $\lambda_l \mapsto
\lambda_l/4p$, $(l=1,\dots,p)$, $p \to \infty$, the eigenvalues in the neighbourhood of $\lambda = 0$ have spacing of order unity, and
moreover exhibit the asymptotic behaviour (\ref{d1}) \cite{Fo94b}.

We therefore see that the hard edge scaled state is specified by the symmetry parameter $\beta$, and the power law exponent $a\beta/2$
(which represents the microscopic repulsion from the origin) in (\ref{AB}). The notation $E_\beta^{\rm hard}(n;(0,s);a\beta/2)$ will be used
to refer to the probability that in the hard edge state so specified, there are exactly
$n$ eigenvalues in the interval $(0,s)$. Similarly, the notation $E_\beta(n;(0,s);{\rm ME}_{\beta,N}(\lambda^{a \beta /2} e^{\beta N \lambda/2}))$ will be
used for the probability that in the ensemble specified by (\ref{AB}), there are $n$ eigenvalues in $(0,s)$.

The most prominent application in physics of the hard edge state is to the study of eigenvalues in lattice QCD. This comes about by the use
of the matrix structure
\begin{equation}\label{AB1}
H = \begin{bmatrix} O_{n \times n} & X \\
X^\dagger & O_{p \times p} \end{bmatrix}
\end{equation}
to model the effective Hamiltonian for given topological anomaly $\nu = n - p$, and with elements of $X$ independent standard Gaussians
which are real $(\beta = 1$), complex $(\beta = 2)$ or real quaternion $(\beta = 4)$. The square of the non-zero eigenvalues of (\ref{AB1})
are equal to the eigenvalues of (\ref{AB}), when the latter corresponds to the eigenvalue PDF of (\ref{AB}).
Significantly, the exact distribution of the $k$-th smallest eigenvalue, $p^{\rm hard}(k;(0,s);\beta a/2)$ say, which is related to
$\{ E_\beta^{\rm hard}(l;(0,s);\beta a/2) \}$ by
$$
p^{\rm hard}(k;(0,s);\beta a/2) = - {d \over ds} \sum_{l=0}^k  E_\beta^{\rm hard}(l;(0,s);\beta a/2),
$$
can be compared against data from lattice QCD simulations \cite{FHKNS09}.

Very recently, the explicit $s \to \infty$ asymptotic form of $  E_\beta^{\rm hard}(n;(0,s);\beta a/2)$ 
for general $\beta > 0$, but requiring $\beta a/2 \in \mathbb Z_{\ge 0}$,
 has been derived  up to an error term which
goes to zero \cite{FS12}. The derivation  relies on an unproved conjecture relating to the asymptotics of a certain
generalized hypergeometric function. Such multivariable special functions, introduced into random matrix theory by
Constantine and Muirhead in the case $\beta = 1$ in the 60's and 70's (see \cite{Mu82}), are finding their way into a
number of recent works relating to asymptotics of eigenvalue distributions \cite{De08,Wa11,Mo11,Fo11y,DL11,Fo12e}.
The result of \cite{FS12} also assumes knowledge of the corresponding asymptotics of 
$E_\beta^{\rm hard}(0;(0,s);\beta a/2)$. The latter was first studied using generalized hypergeometric functions in \cite{Fo94b}, and
the result
\begin{equation}\label{pu}
E_\beta^{\rm hard}(0;(0,s);\beta a/2) \mathop{\sim}\limits_{s \to \infty}
e^{- \beta s/8 + \beta a s^{1/2}/2}
\Big ( {1 \over s} \Big )^{a (\beta a/2 + 1)/4 - \beta a /4}
\tau_{\beta a/2,\beta}^{\rm hard} \Big ( 1 + {\rm O} \Big ( {1 \over s^{1/2}} \Big ) \Big ),
\end{equation}
where
\begin{equation}\label{pp}
\tau_{\beta a/2,\beta}^{\rm hard}= 2^{(1 - \beta/2)a} \Big ( {1 \over 2 \pi} \Big )^{\beta a/4}
\prod_{j=1}^{\beta a /2} \Gamma(2j/\beta),
\end{equation}
was rigorously established for $\beta a/2 \in \mathbb Z_{\ge 0}$, $2/\beta \in \mathbb Z^+$. 
Soon after Dyson's log-Coulomb gas method was used to derive (but not prove) this same asymptotic form, up to the explicit form of the
constant, for general paramaters \cite{CM94}. In the case $\beta = 2$, (\ref{pp}) has been proved for general $\beta a/2 > -1$ in
\cite{DKV11}, and for $|\beta a/2| < 1$ in \cite{Eh10}.
Very recently \cite{RRZ11}, stochastic differential 
equation methods based on tridiagonal matrices realizing the eigenvalue PDF (\ref{AB}) have been used to prove (\ref{pu}) for
general $\beta > 0$, $\beta a/2 > -1$, up to the explicit form of the constant.

These developments motivate us to further consider the asymptotics of the hard edge gap
probability $  E_\beta^{\rm hard}(n;(0,s);\beta a/2)$, both for $n=0$ and general $n \in \mathbb Z^+$.
In Section 2 we summarize previous work on this problem, as has been deduced 
using the method of
generalized hypergeometric functions. In particular, the theory leading to (\ref{pu}) is revised. The aim of Section 3 is to further develop this theory, so
that the form (\ref{pu}) can be rigorously established for $\beta a/2 \in \mathbb Z_{\ge 0}$ and general $\beta > 0$.  In Section 4 we make use of a 
Gaussian fluctuation formula for linear statistics to determine a large deviation formula for
 $E_\beta(n;(0,x);{\rm ME}_{\beta,N}(\lambda^{a \beta /2} e^{\beta N \lambda/2}))$, first in the case $n=0$, then for general $n$.
 Substituting $x = s/(4N)^2$, then taking $N \to \infty$, these large deviation formulas are shown to scale to the asymptotic formulas for
 $  E_\beta^{\rm hard}(n;(0,s);\beta a/2)$ of Section 2, and furthermore give formulas which hold
 for general $\beta a/2$ and $\beta > 0$.
 This latter point requires a suitable continuation of the product in
(\ref{pp}) beyond positive integer values of its upper terminal, which is achieved by making use of the
Barnes double gamma function.
 As a check on these formulas, an asymptotic functional equation known
 from \cite{Fo09}, which relates the asymptotics of the spacing for a particular $(\beta,n,a)$ to
 those for parameters $(4/\beta, \beta(n+1)/2 - 1, \beta (a-2)/2 + 2)$, is shown to hold true.
 
 \section{Generalized hypergeometric function expressions}
 \setcounter{equation}{0}
 It turns out that for $\beta a/2 \in \mathbb Z^+$, the gap probability $E_\beta(n;(0,s);{\rm ME}_{\beta,N}(\lambda^{a \beta /2} e^{\beta N \lambda/2}))$
 can be recognized as an integral representation of certain hypergeometric functions based on Jack polynomials.
 Moreover, the hard edge scaling limit corresponds to a confluence limit of the hypergeometric function in question, giving back
 another hypergeometric function for which there is also a known integral representation, now as a $\beta a/2$--dimensional integral
 and further requiring $2/\beta \in \mathbb Z^+$ \cite{Fo94b}.
 
 The class of generalized hypergeometric functions in question can be defined by the series
 \begin{equation}\label{3.40}
{} _p F_q^{(\alpha)}(a_1,\dots,a_p;b_1,\dots,b_q;x_1,\dots,x_m):=\sum_\kappa \frac{1}{|\kappa|!}\frac{[a_1]^{(\alpha)}_\kappa\dots [a_p]^{(\alpha)}_\kappa }{[b_1]^{(\alpha)}_\kappa
\dots [b_q]^{(\alpha)}_\kappa} 
C_\kappa^{(\alpha)}(x_1,\dots,x_m).
\end{equation}
Here the sum is over all partitions $\kappa_1 \ge \kappa_2 \ge \cdots \ge \kappa_m \ge 0$
of non-negative integers,
$|\kappa| : = \sum_{j=1}^m \kappa_j$, and the generalized Pochhammer symbol $[u]_\kappa^{(\alpha)}$ is defined by
 \begin{equation}\label{C0}
 [a]_\kappa^{(\alpha)} = \prod_{j=1}^m \Big ( a - {1 \over \alpha} (j-1) \Big )_{\kappa_j}, \qquad (a)_k = a (a + 1) \cdots (a+k - 1).
 \end{equation}
The function $C_\kappa^{(\alpha)}(x_1,\dots,x_m)$ is proportional to the Jack symmetric polynomial (see e.g.~\cite[\S 12.6]{Fo10}), and as such
is a homogeneous symmetric polynomial of degree $|\kappa|$. For $m=1$, $C_\kappa^{(\alpha)} = x^{\kappa_1}$ and (\ref{3.40})
reduces to the classical definition of ${}_pF_q$ in one variable. Like ${}_pF_q$ in one variable, ${}{}_pF_q^{(\alpha)}$ exhibits the confluence
property
 \begin{equation}\label{C1}
 \lim_{a_p \to \infty} {}_p F_q^{(\alpha)}\Big (a_1,\dots,a_p;b_1,\dots,b_q; 
 { x_1 \over a_p},\dots, {x_m \over a_p} \Big )= {}_{p-1} F_q^{(\alpha)}(a_1,\dots,a_{p-1};b_1,\dots,b_q;x_1,\dots,x_m). 
\end{equation}

In some cases there are explicit integral formulas for the generalized hypergeometric functions. One example is
\begin{eqnarray}\label{13.44}
\lefteqn{
{}_1 F_1^{(\beta/2)}(-N;a+2m/\beta;t_1,\dots,t_m)  = {1 \over C_{\beta a/2 + m,\beta,m}}  }\nonumber\\
&& \times
\int_0^\infty dx_1 \cdots \int_0^\infty dx_N \,
\prod_{j=1}^N \Big ( x_j^{a \beta /2} e^{-\beta x_j/2} \prod_{k=1}^m (x_j - t_k) \Big ) \prod_{1 \le j < k \le N}
|x_k - x_j|^\beta
\end{eqnarray}
where $C_{\beta a/2 + m,\beta,m}$ is equal to the integral with $t_k=0$, $(k=1,\dots,m)$. This integral formula, obtained explicitly in
\cite{Fo93aa}, follows as a corollary of a similar integral formula for ${}_2 F_1^{(\beta/2)}$ due to Kaneko \cite{Ka93}.

On the other hand the gap probability  $E_\beta(0;(0,s);{\rm ME}_{\beta,N}(\lambda^{a \beta /2} e^{-\beta  \lambda/2}))$ is obtained from
(\ref{AB}) by integrating each $\lambda_l$ over $(0,s)$. Changing variables $\lambda_l \mapsto \lambda_l+s$ allows this
gap probability to be recognized in terms of the integral in (\ref{13.44}), provided $\beta a/2 \in \mathbb Z_{\ge 0}$, thus showing
that in this circumstance \cite{Fo94b}
\begin{equation}\label{E1}
E_\beta(0;(0,s);{\rm ME}_{\beta,N}(\lambda^{a \beta /2} e^{-\beta  \lambda/2})) =
e^{-\beta N s/2} {}_1 F_1^{(\beta/2)}(-N;a;(-s)^{\beta a /2} ).
\end{equation}
Here, in the argument of ${}_1 F_1^{(\beta/2)}$, the notation $(u)^r$ means $u$ repeated $r$ times. Similarly, since by definition
\begin{eqnarray}\label{f2}
\lefteqn{
E_\beta(n;(0,s);{\rm ME}_{\beta,N+n}(\lambda^{a \beta /2} e^{\beta N \lambda/2})) } \nonumber \\
&& = {(N)_n \over n!  } {C_{\beta a/2,\beta,N}  \over C_{\beta a/2,\beta,N+n}}
\int_0^s dy_1 \cdots \int_0^s dy_n \, \Big ( \prod_{l=1}^n y_l^{a \beta /2} e^{-\beta y_l/2} \Big ) \prod_{1 \le j < l \le n}
|y_j - y_l|^\beta \nonumber \\
&& \quad \times
\int_s^\infty dx_1 \cdots \int_s^\infty dx_N \,
\prod_{j=1}^N \Big ( x_j^{a \beta /2} e^{-\beta x_j/2} \prod_{l'=1}^n  |x_l -  y_{l'}|^\beta ) \Big ) \prod_{1 \le j < k \le N}
|x_k - x_j|^\beta, \nonumber \\
\end{eqnarray}
we see from (\ref{13.44}) that for $\beta a/2 \in \mathbb Z_{\ge 0}$, $\beta \in \mathbb Z^+$ \cite{FS12}
\begin{eqnarray}\label{f2a}
\lefteqn{
E_\beta(n;(0,s);{\rm ME}_{\beta,N+n}(\lambda^{a \beta /2} e^{-\beta  \lambda/2})) 
 = {(N)_n \over n!  }
} \nonumber\\
&& \times {C_{\beta a/2,\beta,N} \over C_{\beta a/2,\beta,N+n}} e^{-\beta s (N+k)/2}
\int_0^s dy_1 \cdots \int_0^s dy_n \, \Big ( \prod_{l=1}^n (s - y_l)^{a \beta /2} e^{\beta y_l/2} \Big ) \prod_{1 \le j < l \le n}
|y_j - y_l|^\beta \nonumber \\
&& \quad \times
{}_1 F_1^{(\beta/2)}(-N;a+2n;(-s)^{a \beta/2},(-y_1)^{ a},\dots, (-y_n)^{a}).
\end{eqnarray}

To proceed from (\ref{E1}) and (\ref{f2a}) to the computation of $E_\beta^{\rm hard}(0;(0,s);\beta a/2)$ requires that
\begin{equation}\label{scaling}
s \mapsto s/4N, \qquad N \to \infty.
\end{equation}
Using the confluence (\ref{C1}) it is a simple exercise to deduce from (\ref{E1}) that for $\beta a/2 \in \mathbb Z_{\ge 0}$ \cite{Fo94b}
\begin{equation}\label{f3}
E_\beta^{\rm hard}(0;(0,s);\beta a/2) = e^{-\beta s/8} \,
{}_0 F_1^{(\beta/2)}(a;(s/4)^{\beta a/2}).
\end{equation}
Similarly, using (\ref{C1}) and (\ref{f2a}), together with knowledge of the explicit gamma function functional form of
$C_{a,\beta,N}$ \cite[Prop.~4.7.3]{Fo10} one has that for $\beta a/2 \in \mathbb Z_{\ge 0}$, $\beta \in \mathbb Z^+$
\cite{FS12}
\begin{eqnarray}\label{3.41}
\lefteqn{E^{\rm hard }_{\beta}(n;(0,s);\beta a/2) = A_\beta(n,a)s^{n+(\beta/2)n(n+a-1)} e^{-\beta s/8} } \nonumber \\
&& \times \int_0^1dy_1 \dots \int_0^1dy_n \prod_{j=1}^n (1-y_j)^{\beta a/2} \prod_{1\leq j<k\leq n}|y_k-y_j|^\beta \nonumber \\
&& \times\hspace{1mm} _0F_1^{(\beta/2)}({}_\_;a+2n;(s/4)^{\beta a/2},(sy_1/4)^\beta,\dots,(sy_n/4)^\beta),
\end{eqnarray}
where
\begin{equation} \label{3.42}
A_\beta(n,a)={ 1\over 2^{2n} n!} \Big ( {\beta \over 2} \Big )^n \bigg(\frac{\beta}{4}\bigg)^{n(a+n-1)\beta} \frac{(\Gamma(1+\beta/2))^n}{\prod_{j=0}^{2n-1} \Gamma(a\beta/2+1+j\beta/2)}.
\end{equation}

Next, we want to revise how (\ref{f3}) and also (\ref{3.41})  can be used for purposes of computing the $s \to \infty$ asymptotics.
In relation to (\ref{f3}) this follows from a $\beta a/2$-dimensional  integral formula for ${}_1 F_1^{(\beta/2)}$ in (\ref{f3}).
In addition to the requirement $\beta a/2 \in \mathbb Z_{\ge 0}$, the integral formula also requires that $2/\beta \in \mathbb Z^+$.
Under these assumptions, we then have \cite{Fo94b}
\begin{eqnarray}\label{13.51}
\lefteqn{E^{\rm hard }_{\beta}(0;(0,s);\beta a/2) = B_{a,\beta} e^{-\beta s/8} \Big ( {1 \over s} \Big )^{(-1+2/\beta)\beta a/2}
\Big ( {1 \over 2 \pi} \Big )^{a \beta /2}} \nonumber \\
&& \times
\int_{[-\pi,\pi]^{\beta a/2}} \prod_{j=1}^{\beta a/2} e^{ s^{1/2} \cos \theta_j} e^{i (-1 + 2/\beta) \theta_j}
\prod_{1 \le j < k \le \beta a /2}
|e^{i \theta_k} - e^{i \theta_j} |^{4/\beta} \, d\theta_1 \cdots d \theta_{\beta a/2},
\end{eqnarray}
where
\begin{equation}\label{BB}
B_{a,\beta} = \prod_{j=1}^{a \beta /2} {\Gamma(1 + 2/\beta) \Gamma(2j/\beta) \over \Gamma(1 + 2j/\beta) }.
\end{equation}
It was by using Laplace's asymptotic method to this integral that (\ref{pu}) was derived.
In \cite{Fo94b} the asymptotic form (\ref{pu}) was conjectured to hold true for general $\beta > 0$, and general 
$\beta a /2 \ge -1$ (the latter subject to an appropriate interpretation of the product in (\ref{pp}),
identified in \cite{FS12} as relating to the Barnes double gamma function; see (\ref{feq}) below).

For the generalized hypergeometric function appearing in (\ref{3.41}) there is no explicit integral form analogous to
(\ref{13.51}). Nonetheless a conjectured asymptotic form is available, which states that for
$s \rightarrow \infty$ and $y_1,\dots,y_n \approx 1$ \cite[eq.~(3.45)]{FS12}, 
\begin{eqnarray}\label{gfh}
\lefteqn{{}_0F_1^{(\beta /2)}({}_\_;c;(s/4)^{\beta a/2},(sy_1/4)^\beta,\dots, (sy_n/4)^\beta)} \nonumber \\
 && = {}_0F_1^{(\beta /2)} ({}_\_;c;(s/4)^{\beta (a+2n)/2} ) e^{\beta \sqrt{s}\sum_{j=1}^n(1-y_j)/2 }\Big( 1+{\rm O} \big(\frac{1}{s^{1/2}}\big)\Big),
\end{eqnarray}
and this used in (\ref{3.41})
gives a conjecture for the $s \to \infty$ asymptotics of $E_\beta^{\rm hard}(n;(0,s);\beta a/2)$
\cite[Conj.~8]{FS12}.

\begin{conj}
For $s \to \infty$ we have
\begin{equation}\label{Ef}
{E^{\rm hard }_{\beta}(n;(0,s);\beta a/2) \over E^{\rm hard }_{\beta}(0;(0,s);\beta a/2) } =
\tau_{\beta a/2,\beta}^{\rm hard}(n)
\exp \Big ( - \beta \Big \{ - \sqrt{s} n + \Big ( {n^2 \over 2} + {n a \over 2} \Big ) \log s^{1/2} \Big \} \Big )
\Big ( 1 +  {\rm O}\Big ( {1 \over s^{1/2}} \Big ) \Big ),
\end{equation}
where in the case $\beta n \in \mathbb Z_{\ge 0}$
\begin{equation}\label{2.16a}
\tau_{\beta a/2,\beta}^{\rm hard}(n) =
{2^{-(a+n)\beta n} \over n!}
\Big ( {\beta \over 2} \Big )^{n(a+n-1)\beta/2}
\prod_{j=1}^{\beta n} {\Gamma(a + 2j/\beta) \over (2 \pi)^{1/2}}
{\prod_{j=0}^{n-1} \Gamma(1 + (j+1)\beta/2) \over \prod_{j=n}^{2n-1} \Gamma(1 + (j+a)\beta/2)}.
\end{equation}
\end{conj}

\section{Integral formula for $E_\beta^{\rm hard}(0;(0,s);\beta a/2)$ with $\beta a /2 \in \mathbb Z^+$, $\beta > 0$}
\setcounter{equation}{0}
We would like to generalize (\ref{13.51}) so that the restriction $2/\beta \in \mathbb Z^+$ can be removed.
For this we return to the finite $N$ gap probability formula (\ref{E1}), which has no such restriction.
Thus with $\beta a/2 \in \mathbb Z_{\ge 0}$, but $\beta > 0$ general, we can use a
$\beta a/2$-dimensional integral form of ${}_1 F_1^{(\beta/2)}(-N;a;(-s)^{\beta a /2} )$,
deduced in turn from an integral formula for ${}_2 F_1^{(\beta/2)}(-N,b;c;(s)^{\beta a /2} )$
\cite[Exercises 13.1 Q4(i)]{Fo10} to deduce from (\ref{E1}) the following formula.

\begin{prop}
We have
\begin{eqnarray}\label{f2b}
\lefteqn{
E_\beta(0;(0,s);{\rm ME}_{\beta,N}(\lambda^{a \beta /2} e^{\beta N \lambda/2}))  =  e^{-\beta N s /2}{1 \over M_{\beta a/2}(N,-1+2/\beta,2/\beta)}} \nonumber \\
&& \times
\int_{-1/2}^{1/2} dx_1 \cdots \int_{-1/2}^{1/2} dx_{\beta a/2}  \, \Big ( \prod_{l=1}^{\beta a/2} 
e^{2 \pi i x_l (-1 + 2/\beta)}
(1 + e^{-2 \pi i x_l})^{(-1 + 2/\beta + N)} e^{s e^{2 \pi i x_l}} \Big )
 \nonumber \\
&& \times
\prod_{1 \le j < k \le \beta a/2} |e^{2 \pi i x_k} - e^{2 \pi i x_j}|^{4/\beta} ,
\end{eqnarray}
where $M_n(a,,b,c)$ denotes the Morris integral \cite[(Eq.~4.4)]{Fo10}, which is evaluated
as a product of gamma functions.
\end{prop}

\noindent Proof. \quad
We indeed proceed as in \cite[Exercises 13.1 Q4(i)]{Fo10}. Thus we begin with the formula
\cite[Eq.~(13.11)]{Fo10}
\begin{eqnarray}\label{15.M1aa} 
&&{1 \over M_n(a,b,1/\alpha)} \int_{-1/2}^{1/2}dx_1 \cdots 
\int_{-1/2}^{1/2}dx_n \,
\prod_{l=1}^n e^{\pi i x_l (a-b)}
|1 + e^{2 \pi i x_l}|^{a+b}
(1 + t e^{2 \pi i x_l})^{-r} \nonumber \\&& \qquad \times
\prod_{1 \le j < k \le n} |e^{2 \pi i x_k} - e^{2 \pi i x_j}|^{2/\alpha}
 = {}_2^{} F_1^{(\alpha)}\Big (r,
-b; {1 \over \alpha} (n-1) +a + 1;
(t)^n \Big ). 
\end{eqnarray}
Replacing $t$ by $t/r$ and taking $r \to \infty$ gives
$$
{1 \over M_n(a,b,1/\alpha)}
\int_{-1/2}^{1/2}dx_1 \cdots \int_{-1/2}^{1/2}dx_n \,
\prod_{l=1}^n 
e^{\pi i x_l(a-b)} |1 + e^{2 \pi i x_l}|^{a+b}
e^{-t e^{2 \pi i x_l}} \prod_{1 \le j < k \le n}
|e^{2 \pi i x_k} - e^{2 \pi i x_j}|^{2/\alpha}
$$
$$
= {}_1^{} F^{(\alpha)}_1(-b; a + 1 +  (n - 1) / \alpha;
(t)^n),
$$
where on the RHS use has been made of  (\ref{C1}).  Recalling (\ref{E1}), we obtain
a $\beta a/2$-dimensional integral formula for $E_\beta(0;(0,s);{\rm ME}_{\beta,N}(\lambda^{a \beta /2} e^{\beta N \lambda/2}))$ by
setting $n = \beta a/2$, $\alpha = \beta/2$, $t=-s$, $b=N$ and $a = 2/\beta - 1$. Simple manipulation of
$ |1 + e^{2 \pi i x_l}|^{a+b}$ then gives (\ref{f2b}). \hfill $\square$

\medskip
Essentially the same integral formula (\ref{f2b}), but restricted to $2/\beta \in \mathbb Z^+$, was used in \cite{Fo94b} to compute
the scaled limit (\ref{scaling}) and thus derive (\ref{13.51}). The utility of the assumption $2/\beta \in \mathbb Z^+$ is that the integrand
can readily be rewritten to be analytic in 
\begin{equation}\label{dd}
z_k := e^{2 \pi i \theta_k}, \qquad (k=1,\dots,\beta a/2)
\end{equation}
except for a pole at the origin, thus allowing for the integration domain $|z_k| = 1$, $(k=1,\dots,\beta a/2)$, to be deformed. Deformation of the contours
played a crucial role in the computing of the scaled limit (\ref{scaling}). We will now show that it is possible to deform the contours, and so compute
(\ref{scaling}), without the need to assume $2/\beta \in \mathbb Z^+$.

\begin{prop}
Let $\mathcal{C}$ be the contour which starts at the origin, runs along the negative real axis in the bottom half plane to $z=-1-0i$, then along
a counter clockwise circle to $z = -1 +0i$, and finally back to the original long the negative real axis in the upper half plane.
Let $B_{a,\beta}$ be specified by (\ref{BB}), and assume $\beta a/2 \in \mathbb Z_{\ge 0}$. We have
\begin{eqnarray}\label{13.51x}
\lefteqn{E^{\rm hard }_{\beta}(0;(0,s);\beta a/2) = B_{a,\beta} e^{-\beta s/8} \Big ( {1 \over s} \Big )^{(-1+2/\beta)\beta a/2}
\Big ( {1 \over 2 \pi} \Big )^{a \beta /2}} \nonumber \\
&& \times
\int_{\mathcal C^{\beta a/2}} \prod_{j=1}^{\beta a/2} e^{ s^{1/2} (z_j + 1/z_j)/2} z_j^{(-1 + 2/\beta)}
\prod_{1 \le j < k \le \beta a /2} \Big ( (z_k - z_j) ({1 \over z_k} - {1 \over z_j}) \Big )^{2/\beta} \,
{dz_1 \over 2 \pi i z_1} \cdots {dz_{\beta a/2} \over 2 \pi i z_{\beta a/2}}. \nonumber \\
\end{eqnarray}
\end{prop}

\noindent Proof. \quad
We begin by noting that the integrand in (\ref{f2b}) is symmetric in the variables $\{x_j\}$, so we are free to choose the ordering
\begin{equation}\label{so}
-1/2 \le x_1 < x_2 < \cdots < x_{\beta a/2} < 1/2,
\end{equation}
provided the integral is multiplied by $(\beta a/2)!$. But with this ordering
\begin{equation}\label{O}
\prod_{1 \le j < k \le \beta a/2} |e^{2 \pi i x_k} - e^{2 \pi i x_j}|^{4/\beta} =
\prod_{1 \le j < k \le \beta a/2} \Big ( 2 \sin \pi (x_k - x_j) \Big )^{4/\beta},
\end{equation}
which is an analytic function of each of the variables (\ref{dd}) in the corresponding complex planes cut along the
negative real axis. The factors $\prod_{l=1}^{\beta a/2} 
e^{2 \pi i x_l (-1 + 2/\beta)}
(1 + e^{-2 \pi i x_l})^{(-1 + 2/\beta + N)}$ share this property, which thus becomes a property of the integrand.

Let $\mathcal C_{N,t}$ denote the contour which starts at $z=-1$, runs along the negative real axis in the bottom half plane
to $z=-N/\sqrt{t} - 0 i$, then along a counter clockwise circle to $z = - N/\sqrt{t} + 0i$, and finally back along the negative real
axis in the upper half plane to $z = -1 + 0i$.  The analyticity properties of the integrand just discussed allow us to deform the original
unit circle contours in the variables (\ref{dd}) to the contours $\mathcal C_{N,t}$, provided an ordering equivalent to (\ref{so}) is
adopted. Furthermore, in the variables (\ref{dd}), the integrand and measure in (\ref{f2b}) reads
\begin{eqnarray}\label{Int}
\lefteqn{
\prod_{j=1}^{\beta a/2} e^{ s^{1/2} (z_j + 1/z_j)/2} z_j^{(-1 + 2/\beta)} \Big ( 1 + {1 \over z_j} \Big )^{-1+2/\beta + N}} \nonumber \\
&& \times
\prod_{1 \le j < k \le \beta a /2} \Big ( (z_k - z_j) ({1 \over z_k} - {1 \over z_j}) \Big )^{2/\beta} \,
{dz_1 \over 2 \pi i z_1} \cdots {dz_{\beta a/2} \over 2 \pi i z_{\beta a/2}}.
\end{eqnarray}
This being symmetric in $\{z_j\}$ we can ignore the ordering constraint, provided we remove the earlier introduced factor of
$(\beta a/2)!$.

Now change variables $z_j \mapsto (N/\sqrt{s}) z_j$. We see that in the scaling limit (\ref{scaling}), (\ref{Int}) becomes proportional to
$  ( {1 \over s}  )^{(-1+2/\beta)\beta a/2}$ times the integrand in (\ref{13.51x}), thus accounting for all $s$ dependent terms in 
the latter. The computation of the proportionality relies on also computing the asymptotics of the Morris integral in (\ref{f2b}). But this is
the very same task as already carried out in \cite{Fo94b}, and the factors as presented in (\ref{13.51x}) result. Finally, we note
that with this change of variables and in the $N \to \infty$  limit each contour $\mathcal C_{N,t}$ becomes the contour $\mathcal{C}$,
with the integral being well defined thereon. \hfill $\square$

\medskip
We remark that in the case $2/\beta \in \mathbb Z^+$, the integrand no longer has a branch cut along the negative real
axis. The contributions to the integral of the portions of $\mathcal{C}$ running along the latter therefore cancel, and after
parameterizing the remaining unit circle, we
reclaim (\ref{13.51}).

The computation of the large $s$ asymptotic form of (\ref{13.51x}) is essentially the same as done in \cite{Fo94b} for
(\ref{13.51}). The reason is that method of stationary phase tells us that the maximum contribution to the integrand then comes
from the neigbourhood of $z_j = 1$. Thus the portion of $\mathcal{C}$ running along the negative real axis plays no role in
this limit. The working of \cite{Fo94b} is therefore justified without the restriction $2/\beta \in \mathbb Z^+$, and so we
obtain (\ref{pu}), proved now for $\beta a/2 \in \mathbb Z_{\ge 0}$, and general $\beta > 0$.

\section{Gaussian fluctuation formulas}
\setcounter{equation}{0}
The first aim of this section is to derive the large $N$ asymptotic form of 
$$E_\beta(0;(0,\tilde{s});{\rm ME}_{\beta,N}(\lambda^{a \beta /2} e^{-2 \beta N \lambda})).$$
We will see that by scaling the corresponding asymptotic formula according to 
\begin{equation}\label{ss}
\tilde{s} \mapsto s/(4N)^2, \qquad N \to \infty,
\end{equation}
 we can reclaim (\ref{pu}).
 We will demonstrate too that an analogous procedure can be applied to
 compute the asymptotics of the scaling limit
 for there being $n$ eigenvalues at $\{y_j\}_{j=1,\dots,n}$ in $(0,s)$
in the ensemble ${\rm ME}_{\beta,N+n}(\lambda^{a \beta /2} e^{-2 \beta N \lambda})$.

\subsection{Strategy}
Our method relies on reformulating the problem as one of computing the asymptotic form of a moment of the characteristic polynomial
for ${\rm ME}_{\beta,N}( e^{-2 \beta N \lambda})$. In two recent works by the author \cite{Fo11,Fo12a}, the spectral densities for the
Gaussian, Laguerre and Jacobi $\beta$-ensembles have been similarly formulated, and the corresponding asymptotics computed
by recognizing that such averages can be interpreted as the characteristic function for the linear statistic $V(x) =
\sum_{l=1}^N \log | x - \lambda_l|$. The significance of this is that for ${\rm ME}_{\beta,N}( e^{-2 \beta N \lambda})$, which is the scaled
Laguerre ensemble with $a=0$, eigenvalue density $ \rho_{(1), N}(t)$ supported to leading order on $(0,1)$, it is a known theorem \cite{BG11} that
\begin{equation}
\label{17.1} \left\langle \prod_{l=1}^N (x+\lambda_l)^{c} \right\rangle_{{\rm ME}_{\beta, N}(e^{-2\beta N \lambda})} \mathop{\sim}
\limits_{N \to \infty} e^{c \mu_N (v)} e^{(c \sigma (v))^2/2},
\end{equation}
where, with $v(t):= \log (x+t)$, $ x > 0$
\begin{align}
\label{17.2a} \mu_N(v) &= \int_{0}^{1} \rho_{(1), N}(t) v(t) dt\\
\nonumber (\sigma(v))^2 &= \frac{1}{\beta \pi^2} \int_{0}^{1} dt_1 \frac{v(t_1)} {\big((1-t_1)t_1\big)^{1/2}} \int_{0}^{1} dt_2 \frac{v'(t_2) \big((1-t_2) t_2\big)^{1/2}}{t_2-t_1}  \\
\label{17.2b} &= \frac{1}{2\beta} \sum_{k=1}^{\infty} k a^2_k, \qquad a_k=\frac{2}{\pi}\int_0^{\pi} v\left( \frac{1}{2} + \frac{1}{2}\cos \theta \right) \cos k\theta \; d\theta,
\end{align}
and the notation $\mathop{\sim}
\limits_{N \to \infty} $ means that in the $N \to \infty$ limit the ratio of the LHS and RHS tends to unity.
In words (\ref{17.2a}) says the characteristic function for the linear statistic exhibits Gaussian fluctuations, with an explicit mean (which is of order $N$), and
an explicit variance (which is of order unity).

We will now show proceed to give the details of how this formalism applies to computing the sought large $N$ asymptotic forms.

\subsection{Large $N$ form of $E_\beta(0;(0,\tilde{s});{\rm ME}_{\beta,N}(\lambda^{a \beta /2} e^{-2 \beta N \lambda}))$}
Our first task is to show that $E_\beta(0;(0,\tilde{s});{\rm ME}_{\beta,N}(\lambda^{a \beta /2} e^{-2 \beta N \lambda}))$ can be expressed in terms of
the average on the LHS of (\ref{17.1}).

\begin{lemma}\label{L1}
Let ${\mathcal N}({\rm ME}_{\beta,N}(x^{\beta a/2} e^{-\beta x/2}))$ denote the normalization required to make (\ref{AB}) a probability density function,
and let $\tilde{s} = s/4N$.
We have
\begin{eqnarray}\label{F1}
\lefteqn{E_\beta(0;(0,s);{\rm ME}_{\beta,N}(\lambda^{\beta a/2} e^{- \beta  \lambda/2}))} \nonumber \\
&& = e^{-\beta N s/2} (4N)^{Na}
{{\mathcal N}({\rm ME}_{\beta,N}( e^{-\beta x/2})) \over {\mathcal N}({\rm ME}_{\beta,N}( x^{\beta a/2} e^{-\beta x/2})) }
 \left\langle \prod_{l=1}^N (\tilde{s}+\lambda_l)^{\beta a/2} \right\rangle_{{\rm ME}_{\beta, N}(e^{-2 \beta  N \lambda})}.
\end{eqnarray}
\end{lemma}

\noindent
Proof. \quad 
A simple change of variables $\lambda_l \mapsto \lambda_l + s$ in the definition shows
\begin{eqnarray*}
\lefteqn{
E_\beta(0;(0,s);{\rm ME}_{\beta,N}(\lambda^{\beta a/2} e^{- \beta  \lambda/2})) } \\
&& = e^{-\beta N s/2} {{\mathcal N}({\rm ME}_{\beta,N}( e^{-\beta x/2})) \over {\mathcal N}({\rm ME}_{\beta,N}( x^{\beta a/2} e^{-\beta x/2})) }
 \left\langle \prod_{l=1}^N (s+\lambda_l)^{\beta a/2} \right\rangle_{{\rm ME}_{\beta, N}(e^{- \beta   \lambda/2})}.
 \end{eqnarray*}
We see from the definitions that with $\tilde{s} = s/4N$. The result (\ref{F1}) now follows by noting that the change of variables
$\lambda_l \mapsto 4N \lambda_l$ in the average on the RHS implies
$$
 \left\langle \prod_{l=1}^N (s+\lambda_l)^{\beta a/2} \right\rangle_{{\rm ME}_{\beta, N}(e^{- \beta   \lambda/2}))} = (4N)^{Na}
 \left\langle \prod_{l=1}^N (\tilde{s}+\lambda_l)^{\beta a/2} \right\rangle_{{\rm ME}_{\beta, N}(e^{-2 \beta  N \lambda})}.
$$
\hfill $\square$

\medskip
We know from \cite[eqns.~(6.21) and (6.22)]{FFG06e} and \cite[Eq.~(4.1) with $N=M$]{Fo11} that for ME${}_{\beta,N}(e^{-2 \beta N x})$ the
density on $[0,1]$ is such
\begin{equation}\label{r1}
2 t  \rho_{(1), N}(t^2) = {4 N \over \pi} (1 - t^2)^{1/2} + \Big ( {1 \over2  \beta} - {1 \over 4} \Big )
\Big ( \delta(t - 1) - \delta(t) \Big ) + {\rm O} \Big ( {1 \over N} \Big ).
\end{equation}
With this explicit form, we want to compute (\ref{17.2a}). The following integral evaluation is required.

\begin{lemma}
Let $x> 0$. We have
\begin{equation}\label{r2}
{2 \over \pi} \int_{-1}^1 \log |ix + t| \, (1 - t^2)^{1/2} \, dt= \sqrt{x (x+1)} - x - \log \Big (2 (\sqrt{x^2 + 1} - \sqrt{x} ) \Big ) - {1 \over 2}.
\end{equation}
\end{lemma}

\noindent
Proof. \quad This can be deduced by an appropriate analytic continuation of \cite[Eq.~(3.2)]{Fo11}.
\hfill $\square$

\medskip
\begin{cor}\label{C1c}
We have
\begin{eqnarray}\label{r3}
\lefteqn{
\int_0^1 \log | \tilde{s} + t| \, \rho_{(1),N}(t) \, dt } \nonumber \\
&& = 2N \Big (  \sqrt{\tilde{s} (\tilde{s}+1} - \tilde{s} - \log \Big (2 (\sqrt{\tilde{s}^2 + 1} - \sqrt{\tilde{s}} ) \Big ) - {1 \over 2} \Big )
+  \Big ( {1 \over2  \beta} - {1 \over 4} \Big ) \log {|1 + \tilde{s}| \over |\tilde{s}|}.
 \end{eqnarray}
\end{cor}

\noindent
Proof. \quad This follows by first noting
$$
\int_0^1\log | \tilde{s} + t| \, \rho_{(1),N}(t) \, dt =
\int_{-1}^1 \log | i \sqrt{\tilde{s}} + t| (2t \rho_{(1),N}(t^2)) \, dt,
$$
where on the RHS $2 t  \rho_{(1), N}(t)$ is given by the RHS of (\ref{r1}), extended to be an even function by the
addition of 
$\Big ( \delta(t +1) - \delta(t) \Big )$.  Now (\ref{r3}) can be read off from (\ref{r2}).
\hfill $\square$

\medskip
For the variance as specified by (\ref{17.2b}), a straightforward modification of the working which gave \cite[Eq.~(4.4)]{Fo11}
shows
\begin{equation}\label{r4}
\sigma^2 = - {2 \over \beta} \log \Big ( \tilde{s} ( \tilde{s}+1) \Big )^{1/2} - {4 \over \beta}
\log \Big ( 2 ((\tilde{s} + 1)^{1/2} - \tilde{s}^{1/2}) \Big ).
\end{equation}

Substituting (\ref{r3}) in (\ref{17.2a}) and (\ref{r4}) in (\ref{17.2b}), then substituting into (\ref{17.1}) we read
off the large $N$ asymptotic form of the average on the RHS of (\ref{F1}).

\begin{prop}
We have
\begin{eqnarray}\label{r5}
\lefteqn{ \left\langle \prod_{l=1}^N (\tilde{s}+\lambda_l)^{\beta a/2} \right\rangle_{{\rm ME}_{\beta, N}(e^{-2 \beta  N \lambda})} }\nonumber \\
&& \mathop{\sim}\limits_{N \to \infty}
\exp \Big ( N \beta a \Big ( \sqrt{\tilde{s}(\tilde{s} + 1)} - \tilde{s} -
\log \Big (2 (  (\sqrt{\tilde{s}^2 + 1} - \sqrt{\tilde{s}} ) \Big ) - {1 \over 2} \Big ) \nonumber \\
&& \qquad + {\beta a \over 4} \Big ( {1 \over \beta} - {1 \over 2} \Big ) \log \Big | 1  + {1 \over \tilde{s}} \Big | \Big )  \nonumber  \\
&& \qquad \times \exp \Big ( - {\beta a^2 \over 4} \log ( \tilde{s}(\tilde{s} + 1))^{1/2} +
{\beta a^2 \over 2} \log {(\tilde{s} + 1)^{1/2} + \tilde{s}^{1/2} \over 2} \Big ).
\end{eqnarray}
\end{prop}

Recalling (\ref{F1}), it remains to compute the large $N$ forms of the ratio of normalizations. Each is an example of a limiting
form of the Selberg integral, and so is given by a product of gamma functions (see e.g.~\cite[Prop.~4.7.3]{Fo10}), 
\begin{equation}\label{r6}
{{\mathcal N}({\rm ME}_{\beta,N}( e^{-\beta x/2})) \over {\mathcal N}({\rm ME}_{\beta,N}( x^{\beta a/2} e^{-\beta x/2})) }
=  (\beta / 2)^{N a \beta /2} \prod_{j=0}^{N-1} {\Gamma(1 + j \beta /2) \over \Gamma(a \beta/2 + 1 + j \beta /2) }.  
\end{equation}
In the case that $a \in \mathbb Z_{\ge 0}$, simple manipulation of the product and use of Stirling's
formula shows that the large $N$ form of (\ref{r6}) is
\begin{equation}\label{r6a}
 N^{-aN\beta/2} (N \beta /2)^{-\beta a(a-1)/4} (\pi N \beta)^{-a/2}
 e^{a N \beta /2} \prod_{j=0}^{a-1} \Gamma(1 + j \beta/2).
 \end{equation}
 
 To compute the large $N$ form of (\ref{r6}) without the assumption $a \in \mathbb Z_{\ge 0}$, we
 follow the lead of the recent works \cite{BMS11,Os12} and introduce the Barnes double gamma
 function $\Gamma_2(z;1,\tau)$. This function is related to the usual gamma function through the
 two functional equations
 \begin{equation}\label{feq}
 {1 \over \Gamma_2(z+1;1,\tau)} =
 {\tau^{z/\tau - 1/2} \over \sqrt{2 \pi} } {\Gamma(z/\tau) \over \Gamma_2(z;1,\tau)}, \quad
 {1 \over  \Gamma_2(z+\tau;1,\tau)}  = {1 \over \sqrt{2 \pi}} {\Gamma(z) \over
 \Gamma_2(z;1,\tau)},
 \end{equation}
 and is normalized by requiring that $\lim_{z \to 0} z \Gamma_2(z;1,\tau) = 1$. 
  A result of Shintani \cite{Sh80} gives that $ \Gamma_2(z;1,\tau)$ can be written as an
  infinite product of gamma fuctions,
  \begin{align}\label{fequ}
   \Gamma_2(z;1,\tau) =  & (2 \pi)^{z/2}
   \exp \Big \{ \Big ( {z - z^2 \over 2 \tau} - {z \over 2} \Big ) \log \tau +
   {(z^2 - z) \gamma \over 2 \tau} \Big \} \nonumber \\
   & \times \Gamma(z) \prod_{n=1}^\infty {\Gamma(z+n \tau) \over \Gamma(1 + n \tau)}
   \exp \Big \{ {z - z^2 \over 2n \tau} + (1 - z) \log (n \tau) \Big \},
   \end{align}
   where $\gamma$ denotes Euler's constant.
 Now, as shown
 in \cite{BMS11}, it is easy to check from the underlying recurrence that
 $
 f_{\beta/2}(n+1):= \prod_{j=0}^n \Gamma(1 + \beta j /2)$ can be written in terms of the Barnes double gamma
 function according to 
 \begin{equation}\label{r6b}
 f_{\beta/2}(n+1) = (2 \pi)^{n/2} \tau^{-(n^2 - n(1 - \tau))/2 \tau} {\Gamma(n) \Gamma(1 + n/\tau) \over
 \Gamma_2(n;1,\tau)}, \qquad \tau:= 2/\beta.
 \end{equation}
 Further use of (\ref{feq}) shows that this simplifies to give
  \begin{equation}\label{r6x}
 f_{\beta/2}(n) = (2 \pi)^{(n+1)/2} \tau^{-(n-1)n/2\tau - n/2}
 {1 \over \Gamma_2(n+\tau;1,\tau)}.
 \end{equation}
 But for $a \in \mathbb Z_{\ge 0}$ we have that the product in (\ref{r6a}) is
 equal to $f_{\beta/2}(a) f_{\beta/2}(N) / f_{\beta/2}(N+a)$. This product is uniquely determined by its recurrence in $N$, and
 its value 1 at $N = 0$, and thus the form implied by (\ref{r6b}) persists without the assumption
 $a \in \mathbb Z_{\ge 0}$. Moreoever, the asymptotic form of $\Gamma_2(z;1,\tau)$
 \cite{QHS93} then tells us that the large $N$ form of $f_{\beta/2}(N) / f_{\beta/2}(N+a)$ continues naturally off its
 $a \in \mathbb Z_{\ge 0}$ form, and so can be read off (\ref{r6a}). Thus the latter without
 the assumption $a \in \mathbb Z_{\ge 0}$ reads
 \begin{equation}\label{r6c}
 N^{-aN\beta/2} (N \beta /2)^{-\beta a(a-1)/4} (\pi N \beta)^{-a/2}
 e^{a N \beta /2} f_{\beta/2}(a),
 \end{equation}
 where $f_{\beta/2}(a)$ is specified by (\ref{r6x}).
Substituting (\ref{r5}) and (\ref{r6c}) in (\ref{F1}) 
 with $c= \beta a /2$ the sought large deviation formula results.
 
 \begin{cor}
 Let $\tilde{s} = s/4N$, and let $f_{\beta/2}(a)$ be specified by  (\ref{r6x}).
 With $\tilde{s}$ fixed and $a \in \mathbb Z_{\ge 0}$
we have
\begin{eqnarray}\label{F1a}
\lefteqn{E_\beta(0;(0,s);{\rm ME}_{\beta,N}(\lambda^{\beta a/2} e^{- \beta  \lambda/2}))  \mathop{\sim}\limits_{N \to \infty} } \nonumber \\
&& e^{-\beta 2 N^2 \tilde{s}} (N \beta / 2)^{-\beta a (a - 1) \beta /4} (\pi N \beta)^{-a/2}
f_{\beta / 2}(a) \nonumber \\
&&\times \exp \Big \{ N \beta a \Big ( \sqrt{\tilde{s}(\tilde{s} + 1)} - \tilde{s} +
\log \Big (   \sqrt{\tilde{s}^2 + 1} + \sqrt{\tilde{s}}  \Big ) \Big ) 
+ {\beta a \over 4} \Big ( {1 \over \beta} - {1 \over 2} \Big ) \log \Big | 1  + {1 \over \tilde{s}} \Big | \Big \}  \nonumber \\
&&  \times \exp \Big \{ - {\beta a^2 \over 4} \log ( \tilde{s}(\tilde{s} + 1))^{1/2} +
{\beta a^2 \over 2} \log {(\tilde{s} + 1)^{1/2} + \tilde{s}^{1/2} \over 2} \Big \}.
\end{eqnarray}
\end{cor}

An idea that goes back to Dyson \cite{Dy62a} is to now scale the large deviation formula, which at the hard edge
requires (\ref{scaling}), to deduce the large $s$ expansion of $E_\beta^{\rm hard}(0;(0,s);\beta a/2)$. A rigorous 
justification of this procedure, which plays an essential role in a recent study of asymptotics of the soft edge gap
probability \cite{BEMN10}, requires that a uniform error bound be provided with the large deviation formula \cite{Kr04}.
Unfortunately the present method does not provide us with an error estimate. Nonetheless we find that the hard edge
scaling of (\ref{F1a}) does reclaim (\ref{pu}).

\begin{prop}
Setting $\tilde{s} = s/(4N)^2$, then taking $N \to \infty$, the RHS of (\ref{F1a}) is equal to 
\begin{equation}\label{mg}
A_{a,\beta}\exp \Big ( -{\beta s \over 8} + {\beta a \over 2} \sqrt{s} - {\beta \over 4} a(a-1) \log s^{1/2} - {a \over 4} \log s^{1/2} \Big ),
\end{equation}
where
\begin{equation}
A_{a,\beta} = (\beta / 2)^{-\beta a (a - 1)/4} (\pi \beta)^{-a/2} 2^{-\beta a/2 + a} f_{\beta/2}(a).
\end{equation}
Moreover, in the case $a\beta /2 \in \mathbb Z^+$ we have
\begin{equation}\label{At}
A_{a,\beta} = \tau_{\beta a/2,\beta}^{\rm hard},
\end{equation}
where $\tau_{\beta a/2,\beta}^{\rm hard}$ is given by (\ref{pp}), and thus (\ref{mg}) reclaims the large $s$ asymptotic form of
$E_\beta^{\rm hard}(0;(0,s);\beta a/2)$ as given by (\ref{pu}).
\end{prop}

\noindent
Proof. \quad Obtaining (\ref{mg}) from (\ref{F1a}) is an elementary calculation.
The equality (\ref{At}) can be established by verifying that both sides satisfy the same
defining recurrence, making use of (\ref{feq}) on the LHS.
\hfill $\square$

\subsection{Probability density function for there being $n$ eigenvalues at $\{y_j\}_{j=1,\dots,n}$ in $(0,s)$}
Let $p(y_1,\dots,y_n;(0,s);{\rm ME}_{\beta,N+n}(x^{\beta a/2} e^{-\beta x/2})$ denote the probability density function
that in the ensemble ${\rm ME}_{\beta,N+n}(x^{\beta a/2} e^{-\beta x/2})$
there are $n$ eigenvalues at $\{y_j\}_{j=1,\dots,n}$ in $(0,s)$. The aim of this subsection is to compute the large $N$ form
of this probability density function, with
\begin{equation}\label{s0}
\tilde{s}_0 = {s \over 4N}, \qquad \tilde{s}_j = {(s - y_j) \over 4N} \quad (j=1,\dots,n)
\end{equation}
fixed, and then the scaling limit 
\begin{equation}\label{s0s}
\tilde{s}_0 = s_0/(4N)^2, \qquad  \tilde{s}_j =  {s}_j/(4N)^2  \quad (j=1,\dots,n), \qquad N \to \infty
\end{equation}
of this hard edge form.

\begin{lemma}
In terms of the notation (\ref{s0}) we have
\begin{eqnarray}\label{4.18}
\lefteqn{p(y_1,\dots,y_n;(0,s);{\rm ME}_{\beta,N+n}(x^{\beta a/2} e^{-\beta x/2}))} \nonumber \\
&&
 =
{\mathcal{N}({\rm ME}_{\beta,N}(e^{-\beta x/2})) \over \mathcal{N}({\rm ME}_{\beta,N+n}(x^{a\beta/2}e^{-\beta x/2}))}
{(N)_n \over n!} (4N)^{\beta N (a /2 + n)} e^{-\beta N s/2}\prod_{l=1}^n y_l^{\beta a/2}  \nonumber \\
&&  \times e^{-\beta \sum_{j=1}^n y_j} \prod_{1 \le j < k \le n}|y_j - y_k|^\beta 
\left\langle \prod_{l=1}^N\Big (  (\tilde{s}_0+x_l)^{\beta a/2} \prod_{j=1}^n (\tilde{s}_j + x_l)^\beta  
\Big ) \right\rangle_{{\rm ME}_{\beta, N}(e^{-2 \beta  N x})} 
\end{eqnarray}
\end{lemma}

\noindent
Proof. \quad We essentially follow the strategy of the proof of Lemma \ref{L1}.
\hfill $\square$

\medskip
Comparing the average in (\ref{4.18}) with (\ref{13.44}) we see that in the case $\beta a/2 \in \mathbb Z_{\ge 0}$,
$\beta \in \mathbb Z_{\ge 0}$, we have
\begin{eqnarray}\label{4.18a}
\lefteqn{
{\mathcal{N}({\rm ME}_{\beta,N}(e^{-\beta x/2})) \over \mathcal{N}({\rm ME}_{\beta,N}(x^{(a+2n)\beta/2}e^{-\beta x/2}))}
\left\langle \prod_{l=1}^N\Big (  (\tilde{s}_0+x_l)^{\beta a/2} \prod_{j=1}^n (\tilde{s}_j + x_l)^\beta  
\Big ) \right\rangle_{{\rm ME}_{\beta, N}(e^{-2 \beta  N x})} } \nonumber \\
&& = (4N)^{-N \beta ( a/2 + n )}
{}_1 F_1^{(\beta/2)}(-N;a+2n;(-4N \tilde{s}_0)^{\beta a/2},(-4N \tilde{s}_1)^{\beta},\dots, (-4N\tilde{s}_n)^{\beta}). \nonumber \\
\end{eqnarray}
Furthermore,
we recognise the average as the form given for large $N$ by the RHS of (\ref{17.1}) with $c=1$ and
\begin{equation}\label{vt}
v(t) = {\beta a \over 2} \log (t + \tilde{s}_0) + \beta \sum_{j=1}^n \log ( t + \tilde{s}_j).
\end{equation}
Thus our immediate task is to compute the mean and variance.

\begin{lemma}
With $v(t)$ given by (\ref{vt}), and $\rho_{1,(N)}(t)$ as implied by (\ref{r1}) we have
\begin{eqnarray}\label{4.20}
&&\mu_N(v) = 2N \Big ( {\beta a \over 2} \Big ) \Big (
\sqrt{\tilde{s}_0 (\tilde{s}_0 + 1)} - \tilde{s}_0 - \log \Big ( 2 (\sqrt{\tilde{s}_0^2 + 1} - \sqrt{\tilde{s}_0}) \Big ) - {1 \over 2} \Big ) \nonumber \\
&& \qquad  \qquad  + 2 N \beta \sum_{j=1}^n  \Big (
\sqrt{\tilde{s}_j (\tilde{s}_j + 1)} - \tilde{s}_j - \log \Big ( 2 (\sqrt{\tilde{s}_j^2 + 1} - \sqrt{\tilde{s}_j}) \Big ) - {1 \over 2} \Big ) \nonumber \\
&& \qquad  \qquad  + \Big ( {1 \over 2 \beta} - {1 \over 4} \Big )
\Big \{ {\beta a \over 2} \log \left | {1 + \tilde{s}_0 \over \tilde{s}_0} \right | +
\beta \sum_{j=1}^n \log  \left | {1 + \tilde{s}_j \over \tilde{s}_j} \right |  \Big \}.
\end{eqnarray}
\end{lemma}

\noindent
Proof. \quad We proceed as in the proof of Corollary \ref{C1c}.
\hfill $\square$

\begin{lemma}
Let $\nu_k := - (2 \tilde{s}_k + 1) + 2 (\tilde{s}_k^2 + \tilde{s}_k)^{1/2}$.  We have
\begin{eqnarray}\label{gm}
&& (\sigma(v))^2 = 2 \beta \Big \{ - \Big ( {a \over 2} \Big )^2 \log (1 - \nu_0^2 ) -
\sum_{j=1}^n \log (1 - \nu_j^2) - a \sum_{j=1}^n \log (1 - \nu_0 \nu_j) \nonumber \\
&& \qquad - 2 \sum_{1 \le j_1 < j_2 \le n} \log (1 - \nu_{j_1} \nu_{j_2}) \Big \}.
\end{eqnarray}
\end{lemma}

\noindent
Proof. \quad In the notation of (\ref{17.2b}) the first task is to compute
$$
a_k = {2 \over \pi} \int_0^\pi \Big \{
{\beta a \over 2} \log \Big ( 1 + {\cos \theta \over 2 \tilde{s}_0 + 1} \Big ) +
\beta \sum_{j=1}^n \log \Big ( 1 + {\cos \theta \over 2 \tilde{s}_j + 1} \Big )  \Big \}
\cos k \theta \, d \theta.
$$
According to \cite[Lemma 2]{Fo11}, with $\nu_k$ as specified above
$$
a_k = - {2 \beta \over k} \sum_{j=0}^n w_j \nu_j^k,
$$
where $w_j = a/2$ for $j=0$ and $w_j = 1$ for $j=1,\dots,n$. Squaring this and performing the sum as required in (\ref{17.2b})
gives (\ref{gm}).
\hfill $\square$

\medskip
According to the RHS of (\ref{17.1}) we have that the large $N$ asymptotic form of the average in (\ref{4.18}) is given by
$\exp(\mu_N(v) + (\sigma(v))^2/2)$, with $\mu_N(v)$ specified by (\ref{4.20}) and $ (\sigma(v))^2$ by (\ref{gm}).
We are particularly interested in the double scaling limit obtained by performing (\ref{s0s}) in this asymptotic form
when multiplied by appropriate prefactors.

\begin{prop}
Let the large $N$ asymptotic form of the average  (\ref{4.18}), denoted by annotating the average by $\#$,
be scaled according to (\ref{s0s}).
We have
\begin{eqnarray}\label{4.23a}
&& (4N)^{N \beta ( a/2 + n )}
{\mathcal{N}({\rm ME}_{\beta,N}(e^{-\beta x/2})) \over \mathcal{N}({\rm ME}_{\beta,N}(x^{(a+2n)\beta/2}e^{-\beta x/2}))}
\left\langle \prod_{l=1}^N\Big (  (\tilde{s}_0+x_l)^{\beta a/2} \prod_{j=1}^n (\tilde{s}_j + x_l)^\beta  
\Big ) \right\rangle_{{\rm ME}_{\beta, N}(e^{-2 \beta  N x})}^{\#}
\nonumber \\
&&  \sim
\tilde{A}_{a,n,\beta} e^{(\beta a/2) \sqrt{s_0} + \beta \sum_{j=1}^n \sqrt{s_j}} 
\exp\Big \{  - {1 \over 2} \Big ( 1  - {\beta \over 2} \Big ) \Big (a \log {s_0 \over 4} + 2 \sum_{j=1}^n \log {s_j \over 4} \Big ) \Big \}  
\nonumber \\
&& \quad \times
\exp\Big ( - \beta \Big \{
\Big ( {a \over 2} \Big )^2 \log \sqrt{s_0}  + \sum_{j=1}^n \log \sqrt{s_j}  +
a \sum_{j=1}^n \log \Big ( {\sqrt{s_0} + \sqrt{s_j} \over 2} \Big )  \Big \} \Big )
\nonumber \\
&&
\quad \times \exp \Big ( 
-2  \beta \sum_{1 \le j_1 < j_2 \le n} \log \Big ( {\sqrt{s_{j_1}} + \sqrt{s_{j_2}} \over 2} \Big )\Big ),
\end{eqnarray}
where, with $f_{\beta/2}$ specified by (\ref{r6x}),
\begin{equation}
\tilde{A}_{a,n,\beta} = (\beta/2)^{-(\beta/4)(a+2n-1)(a+2n)} (\pi \beta)^{-(2n+a)/2}
f_{\beta/2}(a+2n-1).
\end{equation}
\end{prop}

According to (\ref{4.18a}) and (\ref{C1}), the scaled limit (\ref{s0s}) of the LHS of (\ref{4.23a})
is equal to
\begin{equation}\label{x1}
{}_0 F_1^{(\beta / 2)}({}_\_;a+2n;({{s}_0 \over 4})^{\beta a/2},({{s}_1\over 4})^{\beta},\dots, ({{s}_n \over 4})^{\beta}).
\end{equation}
Thus we obtain, as a conjecture, the corresponding large $\{s_j\}_{j=0,\dots,n}$ asymptotic form.

\begin{conj}\label{C2d}
For large values of the arguments $\{s_j\}_{j=0,\dots,n}$, the generalized hypergeometric function (\ref{x1})
has the asymptotic form given by the RHS of (\ref{4.23a}), up to terms which vanish as the arguments
approach infinity.
\end{conj}

In the case $\beta = 4$ the asymptotic expansion of (\ref{x1}) can be derived from a matrix integral
representation not available for general parameters \cite{Mu78}.  The expression implied
by the result of \cite{Mu78} is in precise agreement with the conjecture. Furthermore, this
asymptotic form is also consistent with the conjecture (\ref{gfh}).

In keeping with the origin of (\ref{x1}), and with the scaled variables as specified by (\ref{s0}) and (\ref{s0s}),
we can deduce from (\ref{4.18}) that
\begin{eqnarray*}
\lefteqn{s^n p(sy_1,\dots,sy_n;(0,s);{\rm ME}_{\beta,N+n}(x^{\beta a/2} e^{- \beta x/2}))} \nonumber \\ &&
\sim A_\beta(n,a) s^{n+(\beta/2)n(n+a-1)} e^{-\beta s/8}
\prod_{j=1}^n (1-y_j)^{\beta a/2} \prod_{1 \le j < k \le n} | y_k - y_j |^\beta
 \nonumber \\ && \qquad \times
{}_0 F_1^{(\beta / 2)}({}_\_;a+2n;({{s} \over 4})^{\beta a/2},({{s}(1-y_1)\over 4})^{\beta},\dots, ({{s}(1-y_n) \over 4})^{\beta}).
\end{eqnarray*}
But 
\begin{eqnarray*}
&&E_\beta(n;(0,s);{\rm ME}_{\beta,N+n}(\lambda^{a \beta /2} e^{\beta N \lambda/2})) \nonumber \\
&& \qquad
\quad = s^n \int_0^1 dy_1 \cdots \int_0^1 dy_n \,
 p(sy_1,\dots,sy_n;(0,s);{\rm ME}_{\beta,N+n}(x^{\beta a/2} e^{- \beta x/2}))
 \end{eqnarray*}
 and thus, recalling the consistency of Conjecture \ref{C2d} with (\ref{gfh}), we reclaim
 (\ref{Ef}), but now with requirement that $\beta n \in \mathbb Z_{\ge 0}$ relaxed by writing
 \begin{align}\label{a1}
 \prod_{j=1}^{\beta n}{  \Gamma(a + 2j/\beta) \over (2 \pi)^{1 /2} }& = 
 {\Gamma_2(a+2/\beta;1,2/\beta) \over
 \Gamma_2(2n + a + 2/\beta;1,2/\beta)} \nonumber \\
 & ={\tau^{2n^2/\tau + 2n a /\tau - n/\tau + n} \over (2 \pi)^n}
 {f_{1/\tau}(2n+a) \over f_{1/\tau}(a)}, \qquad \tau:= 2/\beta.
 \end{align}
 
 A significant check on (\ref{Ef}) with the substitution (\ref{a1}),
 supplemented by (\ref{pu}) and (\ref{At}), is to verify that it satisfies the asymptotic
 functional equation \cite{Fo09}
  \begin{equation}\label{fa2}
 E_\beta^{\rm hard}(n;(0,s/\tilde{s}_\beta);\beta a/2) 
 \mathop{\sim}\limits_{s \to \infty}
 E_{4/\beta}^{\rm hard}( \beta (n+1)/2 - 1; (0,s/\tilde{s}_{4/\beta});
 a - 2 + 4/\beta),
 \end{equation}
 where $\tilde{s}_{4/\beta} (\beta / 2)^2 = \tilde{s}_\beta$. Note that on the RHS the value
 for the number of particles in the gap is $\beta (n+1)/2 - 1$ and thus not necessarily an
 integer. In addition to (\ref{a1}) we should also rewrite the remaining products in
 (\ref{2.16a}) according to
 \begin{equation}\label{a2}
{\prod_{j=0}^{n-1} \Gamma(1 + (j+1)\beta/2) \over \prod_{j=n}^{2n-1} \Gamma(1 + (j+a)\beta/2)}
= {f_{\beta/2}(n+1) f_{\beta/2}(n+a) \over f_{\beta/2}(2n+a)},
\end{equation}
where $f_{\beta/2}(n)$ is given by (\ref{r6x}). Combining with (\ref{a1}) then gives
 \begin{align}\label{a2x}
 \tau_{\beta a/2,\beta}^{\rm hard}(n) & =
 {2^{-(a+n)2n/\tau} \over n!} {\tau^{n(a+n)/\tau + n} \over (2\pi)^n}
 {f_{1/\tau}(n+1) f_{1/\tau}(n+a) \over f_{1/\tau}(a)}\nonumber \\
 & =  {2^{-(a+n)2n/\tau} \over \Gamma((n+1)/\tau)}  {\tau^{n(a+n)/\tau -1/2} \over
 (2 \pi)^{n +(1-\tau)/2}} {f_{1/\tau}(n+2-\tau) f_{1/\tau}(n+a) \over f_{1/\tau}(a)}, \quad \tau=2/\beta, 
 \end{align}
 where the second line follows from use of the definition (\ref{r6x}) and the functional
 properties (\ref{feq}).

We showed in \cite{FW11} that (\ref{fa2}) is satisfied up to the constant term in the corresponding
asymptotic expansions of both sides. With $\tilde{s}_{4/\beta}=1$, the constant term on the LHS
of (\ref{fa2}) is
\begin{equation}\label{r1}
  \tau_{\beta a/2,\beta}^{\rm hard}  \, \tau_{\beta a/2,\beta}^{\rm hard}(n)
(\beta/2)^{a(\beta a/2 + 1)/2 - \beta a /2 +(n^2+na)\beta/2},
\end{equation}
while on the RHS the constant term is
\begin{equation}\label{r2}
\tau_{a-2+4/\beta,4/\beta}^{\rm hard} \, \tau_{a-2+4/\beta,4/\beta}^{\rm hard}(\beta (n+1)/2 - 1) .
\end{equation}
Use of (\ref{a2x}) and (\ref{At}) shows that (\ref{r1}) is equal to
\begin{equation}\label{rs1}
 {2^{-(a+n)2n/\tau-a/\tau + a} \over  \Gamma((n+1)/\tau)} 
 {\tau^{ -1/2} \over (2\pi)^{n+a/2+(1-\tau)/2}} f_{1/\tau}(n+1-\tau) f_{1/\tau}(n+a) .
 \end{equation}
 
 To verify the equality between (\ref{r2}) and (\ref{rs1}),
we require the fact that the Barnes double gamma function has the inversion property
\cite{KO98}
\begin{equation}\label{mm}
\Gamma_2(n;1,\tau) = 
\tau^{-(1 + n^2/2 \tau) + n (1 + \tau)/2 \tau } \Gamma_2({n \over \tau};1,{1 \over \tau}).
\end{equation}
Recalling (\ref{r6b}) this implies
\begin{equation}\label{mm1}
f_{1/\tau}(a) = {(2 \pi)^{(a-1)/2} \over (2 \pi)^{(a-1)/2\tau}}
{\tau^{(1-a)/2} \over \tau^{a(a-1)/2\tau}} f_\tau({a-1 \over \tau} + 1).
\end{equation}
Making use of this in (\ref{r2}) shows the latter is equal to
\begin{equation}\label{mm2}
{t^{-a'(a'-1)\tau/2 -(a'/2 + n') - n'(a'+n') t} 2^{-(a'+n')2n't - a't + a'} \over
\Gamma(n'+1) (2 \pi)^{n'+a'/2}} f_t(n'+1) f_t(n'+a'),
\end{equation}
where $a'= (a-2)t + 2$, $n' = (n+1)t - 1$ and $t=1/\tau =\beta/2$. The expressions
(\ref{a2x}) and (\ref{At}) reveal that this is precisely the RHS of (\ref{r2}), thus verifying
that the functional equation (\ref{fa2}) is a property of the asymptotic expansion of
$E_\beta^{\rm hard}(n;(0,s);\beta a/2)$ as implied by (\ref{pu}), (\ref{Ef}), 
(\ref{a2x}) and (\ref{At}).

\subsection*{Acknowledgement}
This work was supported by the Australian Research council.


\providecommand{\bysame}{\leavevmode\hbox to3em{\hrulefill}\thinspace}
\providecommand{\MR}{\relax\ifhmode\unskip\space\fi MR }
\providecommand{\MRhref}[2]{%
  \href{http://www.ams.org/mathscinet-getitem?mr=#1}{#2}
}
\providecommand{\href}[2]{#2}

\end{document}